# Phonon Softening and Pressure-Induced Phase Transitions in Quartz Structured Compound, FePO$_4$

R. Mittal[1,2], R. P. Hermann[3,4], M. Angst[3], S. L. Chaplot[2], E. E. Alp[5], J. Zhao[5], W. Sturhahn[5] and F. Hatert[6]

[1]*Juelich Centre for Neutron Science, IFF, Forschungszentrum Juelich, Outstation at FRM II, Lichtenbergstr. 1, D-85747 Garching, Germany*
[2]*Solid State Physics Division, Bhabha Atomic Research Centre, Mumbai 400 085, India*
[3]*Institut für Festkörperforschung, JCNS and JARA-FIT, Forschungszentrum Jülich, D-52425 Jülich, Germany*
[4]*Département de Physique, B5, Université de Liège, B-4000 Sart-Tilman, Belgium*
[5]*Advanced Photon Source, Argonne National Laboratory, Argonne, IL 60439, USA*
[6]*Laboratoire de Minéralogie, B18, Université de Liège, B-4000 Sart-Tilman, Belgium*

Nuclear resonant inelastic x-ray scattering on quartz structured $^{57}$FePO$_4$ as a function of pressure, up to 8 GPa reveals hardening of the low-energy phonons under applied pressures up to 1.5 GPa, followed by a large softening at 1.8 GPa upon approaching the phase transition pressure of ~2 GPa. The pressure-induced phase transitions in quartz-structured compounds have been predicted to be related to a soft phonon mode at the Brillouin-zone boundary (1/3, 1/3, 0) and to the break-down of the Born-stability criteria. Our results provide the first experimental evidence of this predicted phonon softening.

Minerals of the silica family are amongst the most abundant minerals in the earth's crust and have many commercial applications. α-quartz is the most stable polymorph of silica at ambient conditions. The high pressure phase transformations of α-quartz [1-7] have considerable fundamental interest in condensed matter physics, high-pressure physics, and geophysics. α-quartz is considered a model system for studies of phase transitions, lattice dynamics and chemical bonding in tetrahedral framework structures. The pressure-induced amorphization and the transition to an intermediate crystalline phase in quartz or quartz structured compounds have been studied in detail [1-21] by high pressure experiments and computer simulation based on semiempirical potentials as well as *ab-initio* methods. High pressure calculations [16-21] predict the softening of a phonon mode at the Brillouin-zone boundary point (1/3, 1/3, 0) and the break-down of the Born-stability criteria [22], *i.e.* the requirement of a positive definite elastic tensor, predictions that have been associated with the observed phase transitions. However, experimental evidence of the phonon softening is lacking so far, because the transition pressure for quartz is rather high, ~20 GPa, and the experiments to study the phonon softening by inelastic neutron or x-rays scattering at such high pressures are rather difficult. It is thus interesting to investigate other quartz-structured compounds that exhibit this transition at lower pressures.

FePO$_4$ is homeotypic with the SiO$_2$ quartz structure and is obtained by replacing half the Si by Fe and the other half by P. The structure consists of corner-linked FeO$_4$ and PO$_4$ tetrahedral units. FePO$_4$ undergoes an irreversible transformation [14] to coexisting crystalline orthorhombic, with space group *Cmcm*, and amorphous phases at a pressure of ~2.5 GPa. Upon this phase transition, the Fe oxygen coordination is modified from tetrahedral to octahedral. Calculations indicate phonon softening at (1/3, 1/3, 0) in the FePO$_4$ quartz phase at ~2 GPa, and that the low energy phonons mainly involve the Fe atoms [19].

Herein we report on the partial vibrational density of states of Fe in $^{57}$FePO$_4$ as a function of pressure up to 8 GPa using nuclear resonant inelastic X-ray scattering (NRIXS) [23,24]. We observe phonon hardening up to 1.5 GPa followed by large softening at 1.8 GPa as the phase transition pressure of ~2 GPa is approached. We are providing the first experimental verification of the microscopic origin of the pressure-induced transformations in quartz structured compounds, which are related to the softening of a phonon mode at the Brillouin-zone boundary (1/3, 1/3, 0) and break-down of the Born-stability criteria**.**

The powder sample of $^{57}$FePO$_4$, prepared from a stoichiometric mixture of NH$_4$H$_2$PO$_4$ and $^{57}$Fe$_2$O$_3$ (95% enrichment) heated in air at 950$^o$C during two days. Phase purity was checked by Mössbauer spectroscopy. The spectrum at ambient temperature and pressure is a doublet with an isomer shift relative to α-iron of 0.261(2) mm/s and a quadrupole splitting of 0.631(3) mm/s, with no trace of impurity. The NRIXS spectra were recorded on the 3ID beam line of the Advanced Photon Source, Argonne National Laboratory. The sample was loaded in a beryllium gasket with a central hole of diameter of 100 µm and thickness 50 µm. A pair of diamond anvils with culet size of 500 µm was used in the diamond anvil cell. The x-ray beam was micro focused to 10 µm in diameter. A methanol–ethanol (4:1) mixture was used as pressure transmitting medium and the pressure was determined using the ruby fluorescence line. The measurements were performed at ambient pressure (100 kPa), and at 0.2, 1.5, 1.8, 2.1, 2.6, 5, and 8 GPa, all at room temperature. X-ray diffraction patterns were also collected at these pressures with the wavelength of 0.8602 Å up to 25$^o$ in 2θ using a MAR345 image plate detector.

The $^{57}$Fe nuclei in the sample were excited by the 14.4125-keV x-ray beam and three avalanche photodiodes (APD) were used to collect the NRIXS signals. A fourth APD in forward direction was used either to measure the resolution function, ~1.2 meV full width at half maximum, during the NRIXS experiment using a $^{57}$Fe metal foil or to measure the nuclear forward scattering (NFS). The measured inelastic spectra, *i.e.* count rate as a function of the energy difference between the incident photon energy and the nuclear transition energy, were converted to partial phonon density of states as described in Ref. [25] using the PHOENIX code, available at APS.

In order to monitor the phase transitions upon pressurizing the sample, we have recorded x-ray diffraction and nuclear forward scattering. The X-ray diffraction patterns [Fig. 1(a)] are in qualitative agreement with previous work [14] and reveal the onset of the trigonal to orthorhombic phase transition at 2.1 GPa, a pressure at which the trigonal and orthorhombic phases coexist. At 2.6 GPa the contribution from the trigonal phase is reduced but still discernable, at higher pressures the diffraction patterns correspond to the orthorhombic phase, with likely coexistence of an amorphous phase. The information about the long range order extracted from the diffraction data was complemented with nuclear forward scattering [Fig. 1(b)], a local probe that reveals the change in the Fe coordination at the phase transition: Up to 1.8 GPa, the spectra are well described with a single component of the low-pressure trigonal phase, with a quadrupole splitting $\Delta E_Q \sim 0.75$ mm/s. At pressures above 2.6 GPa the spectra are fitted with a single component that corresponds to the high-pressure orthorhombic phase, with a larger quadrupole splitting [26], decreasing from 2.0 to 1.5 mm/s upon increasing pressure [inset of Fig. 1(b)]. At 2.6 GPa, the majority component of the spectrum (~90.2 %) exhibits the large quadrupole splitting of the orthorhombic phase. At 2.1 GPa, the diffraction data reveals the onset of the trigonal to orthorhombic phase transition. However, only a poor fit of the NFS data could be obtained, with a majority component that exhibits the low pressure phase quadrupole splitting ($\Delta E_Q = 0.69$ mm/s). The x-ray diffraction and NFS data are in perfect agreement, except at 2.1 GPa, where differences may be related to sensitivity of these methods to long and short range order, respectively. Importantly, both methods indicate without doubt that at 1.8 GPa no significant fraction of the sample has undergone the phase transition, and thus that at this pressure the sample is in the trigonal phase. Note that the NFS spectra reported in Fig. 1(b) have been measured after measuring the nuclear inelastic scattering in order to rule out any creeping effect on the pressure during the measurements.

The pressure-dependence of the partial density of states of Fe in $^{57}$FePO$_4$ [Fig. 2] reveals gradual stiffening of the lattice up to 1.5 GPa, softening at 1.8 GPa, and the emergence of new phonon modes, characteristic of the orthorhombic phase, at 2.1 GPa and higher pressures. The observed phonon spectrum at ambient pressure is in fair agreement with our calculations (bottom of Fig. 2) [19]. The observations of hardening of the low energy modes below 8 meV upon increasing the pressure up to 1.5 GPa and the subsequent softening at 1.8 GPa is in qualitative agreement (Fig. 3a) with our lattice dynamical calculations (Fig. 3b) [19] although in the calculations the softening starts at 2.5 GPa and the first peak in the calculation is predicted at a lower energy than that in the experiment. These calculations also suggest [19] that a softening of phonon mode at (1/3, 1/3, 0) (Fig. 3c) is responsible for overall softening of phonons below 8 meV.

At higher energies, the changes in the phonon spectra (Fig. 2) arise from the variation of unit cell volume with pressure (as indicated by the diffraction data, Fig. 1a), as well as the trigonal to orthorhombic phase transition at ~2.1 GPa. As pressure is increased the intensity of all the peaks except two peaks at ~16 and ~21 meV decreases across the trigonal to orthorhombic phase transition, while these two low intensity peaks exhibit a large intensity increase. The phonon peaks at 8, 29, and 45 meV are unique to the low pressure phase at and below 1.8 GPa, while the peaks at 16 and 21 meV are unique to the high pressure phase at and above 2.6 GPa. At 2.1 GPa all of these peaks are present, indicating phase coexistence, in agreement with the diffraction data (Fig. 1(a)). At 2.6 GPa, the minor contribution from the trigonal phase is not resolved.

The average velocity of sound has been derived (Fig. 4(a)) from the parabolic fit to the low energy part of the partial density of states (Fig. 2) at various pressures. The decrease in the sound velocity (Fig. 4(a)) at 1.8 GPa and the subsequent increase at 2.6 GPa are consistent with the decrease in the acoustic phonon energies at 1.8 GPa and the subsequent increase in the orthorhombic phase as shown in Figs. 2 and 3(a). We note that this softening of zone centre acoustic phonon modes (see also Fig. 4(a) inset) has not been predicted by the calculations of Ref. [19] and is related to incipient break-down of the Born-stability criteria [27]**.**

Further evidence of the observed softening at 1.8 GPa is obtained directly from the measured inelastic spectra by using Lipkin's sum rule [28] in order to extract the mean-square displacement ($<u^2>$) of Fe independently of the extraction of the phonon density of states. The recoil fraction of nuclear absorption (1-$f_{LM}$) was calculated from the area of the inelastic part of the normalized spectrum and $<u^2>$ has been obtained from the Lamb-Mössbauer factor ($f_{LM}$) via $f_{LM} = \exp(-k^2 \cdot <u^2>)$, where k=7.31 Å$^{-1}$ corresponds to the photon energy of 14.4 keV. The main feature visible in the variation of $<u^2>$ with pressure (Fig. 4(b)) is a ~20 % increase from 1.5 GPa to 1.8 GPa, revealing the larger Fe atomic displacements due to the softening of the Fe phonon modes. At higher pressures we observe smaller $<u^2>$, as expected, because the Fe atoms are octahedrally bonded in the orthorhombic phase as opposed to their tetrahedral coordination in the trigonal phase. The softening at 1.8 GPa, just below the trigonal to orthorhombic phase transition, is thus established both by the indirect extraction of the DOS and the speed of sound form the NRIXS and by the more direct consideration of the atomic displacements.

High pressure experiments on α-quartz have shown [3] an intermediate crystalline phase before amorphization, the transition to which according to calculations [21] involves unit-cell multiplication and is associated with the phonon softening at the (1/3, 1/3, 0) point in the Brillouin zone. Following this softening at the (1/3, 1/3, 0) point, nearly simultaneous softening is predicted in the whole branch of transverse acoustic phonons from the zone centre to (1/3, 1/3, 0) [27]. The pressure-induced amorphization in alpha quartz [21] and ice-XI [29] is believed to be due to this whole-branch softening. In FePO$_4$, pressure-induced amorphization has been observed [14] to accompany the transition to the *Cmcm* phase. Our observation in FePO$_4$ of an increase in the density-of-states at very low energies, a decrease of the speed of sound, and an increase of the average Fe atom displacement at a pressure slightly below the trigonal to orthorhombic phase transition, provides the first experimental support for this scenario in a quartz-structured compound.

In conclusion, we have presented a study of the Fe partial phonon density of states in FePO4 by nuclear resonant inelastic scattering, which reveals a large softening of low-energy phonons as a precursor to the pressure-induced phase transition. While phonon-softening scenarios have long been proposed to explain pressure induced phase transitions in quartz-structured compounds, we provided the first direct experimental support for that scenario in a compound of this family.

**Acknowledgement**

We thank Dr. M. Lerche, HP-Sync, APS, for his support during the experiment and Dr. M. Sougrati for help during acquisition of the Mössbauer data. Use of the Advanced Photon Source is supported by the U.S. Department of Energy, Basic Energy Science, Office of Science, under Contract No. W-31-109-ENG-38. FH acknowledges the FRS-F.N.R.S. for his position as research associate.

FIG. 1. (Color online) (a) Pressure dependent x-ray diffraction patterns of $FePO_4$ with the calculated patterns for the trigonal (ambient) and orthorhombic phase (8 GPa) shown at the bottom and top, respectively. (b) Pressure dependent nuclear forward scattering by $^{57}FePO_4$ (successive spectra are offset by three orders of magnitude for clarity). The symbols and lines correspond to the data and the fit, respectively. Inset: corresponding quadrupole splitting.

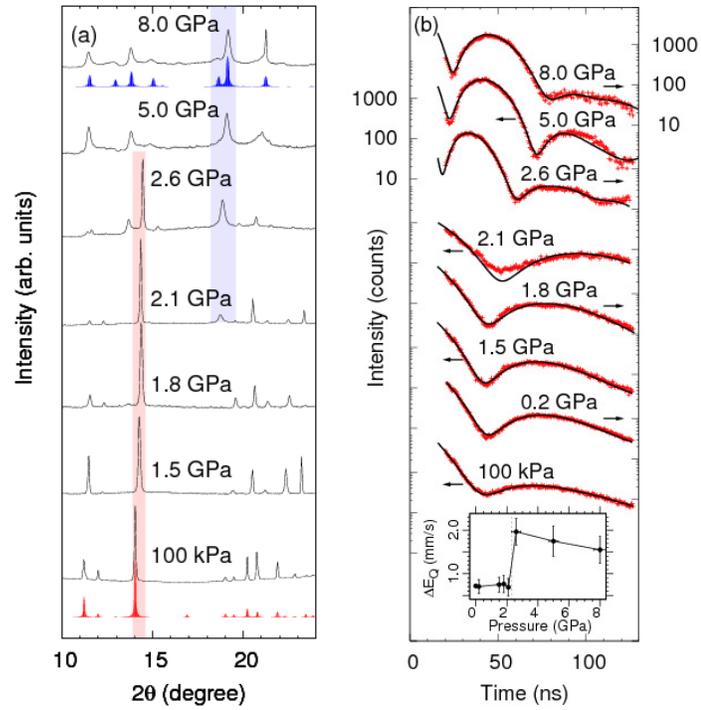

FIG. 2 (Color online) Pressure dependent experimental iron partial density of states (PDOS). Successive spectra are offset by 0.03 meV$^{-1}$. The calculated PDOS in the quartz phase at ambient pressure (convoluted with a Gaussian of FWHM of 1.25 meV) is shown at the bottom. For clarity, only every 4$^{th}$ error bar is shown.

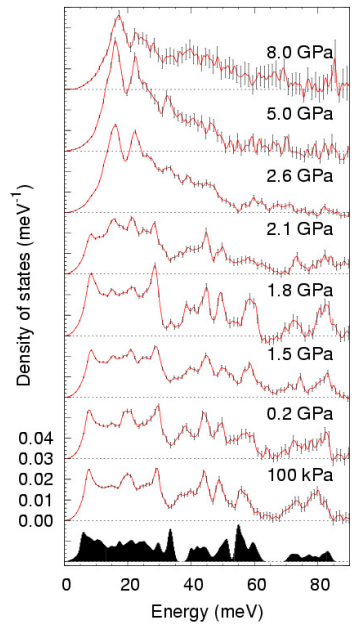

FIG. 3 (Color online) Detail of experimental (a) and calculated (b) [19] iron partial density of states in FePO$_4$ between 0 and 12 meV at selected pressures. (c) Calculated [19] pressure dependence of a transverse acoustic branch along (110).

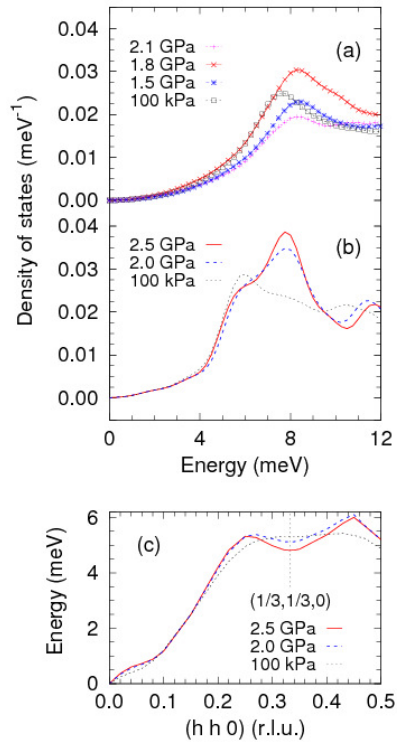

FIG. 4. (Color online) Experimental pressure dependence of the sound velocity (a) and the mean-square displacements of Fe (b). Inset: pressure dependence of the reduced PDOS, $g(E)/E^2$, at selected pressures. At low energies $g(E)/E^2$ is inversely proportional to the cube of the average speed of sound [30].

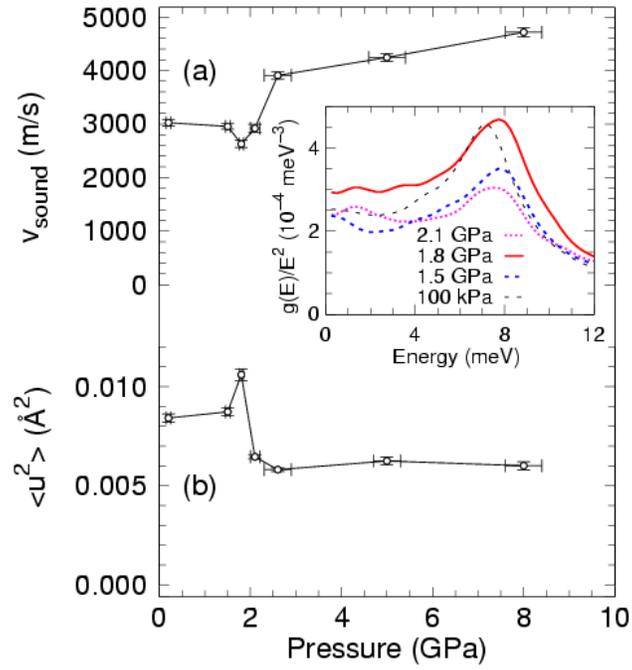